\documentclass[showpacs,twocolumn,prd,aps]{revtex4}
\usepackage{amsfonts}
\usepackage{amsmath}
\usepackage{amsbsy}
\usepackage{amssymb}
\usepackage{epsfig}
\input{epsf}

\begin{document}

\title{Charged black hole accelerated by spatially homogeneous electric\\ field
of Bertotti-Robinson (AdS${}^2\times \mathbb{S}^2$) space-time
}

\author{G.A.~Alekseev}
 \email{G.A.Alekseev@mi-ras.ru}
\affiliation{\centerline{
\hbox{Steklov Mathematical
Institute of the Russian Academy of Sciences,}}\\
\centerline{\hbox{Gubkina str. 8, 119991, Moscow, Russia}}
}

\begin{abstract}
\noindent
A simple exact solution of the Einstein - Maxwell field equations for charged non-rotating black hole accelerated by an external electric field is presented. The background space-time, described by the well known Bertotti-Robinson solution, contains a spatially homogeneous electric field and possess the topology  AdS${}^2\times \mathbb{S}^2$. The black hole mass $m$, its charge $e$ and the value of the background electric field $E$ are free parameters of the constructed solution. In the ``rigid'' (non-inertial) reference frame comoving a black hole, this solution is static. The value of  acceleration of a black hole due to its interaction with the external electric field is determined by the condition of vanishing of conical singularities on the axis of symmetry. The dynamics of a charged black hole in the external electric field  is compared with the behaviour of a charged test particle with the same charge to mass ratio.

\end{abstract}


\maketitle

\section*{Introduction and results}
The most part of our present knowledge of different aspects of interaction of black holes with the external fields was obtained using various perturbation methods. For investigation of such phenomena in the strong field regimes, the only alternative to numerical calculations is the addressing to exact solutions which could describe some (although idealized) models and particular forms of these interactions. However, among a great variety of exact solutions of the Einstein's field equations which we know today,  there is a very small number of solutions describing the interaction of  black holes with the fields of the external sources \cite{SKMHH:2003}, \cite{Griffiths-Podolsky:2009} and, very few of these solutions possess a \emph{dynamical character}.

The solutions for black holes interacting with external magnetic fields arose from the Ernst's observation \cite{Ernst:1976a} that a Harrison-like transformation \cite{Harrison:1968} of stationary axisymmetric vacuum or electrovacuum solution allows to immerse its source into the Melvin magnetic universe and to obtain various ``magnetized'' black hole solutions \cite{Ernst-Wild:1976}.

The well known ``C-metrics'' describe uniformly accelerated black holes which acceleration is caused by ``nodal'' singularities (strings or struts with conical points) located on the symmetry axis. The hystory of discoveries and various properties of these solutions were described in \cite{Kinnersley-Walker:1970}, \cite{Ashtekar-Dray:1981}.
The``generalized C-metric'' was constructed by Ernst in \cite{Ernst:1976b} where the Harrison-like transformation was used to obtain the solution for the Reissner-Nordstr$\ddot{o}$m black hole immersed into \emph{electric} Melvin universe background. It was shown there that for appropriate values of the external electric field and black hole parameters, all nodal singularities disappear and the black hole acceleration is determined  by its interaction with the background electric field. For the rotating case, the similar results were obtained by Bicak and Kofron \cite{Bicak-Kofron:2010}.

However, in Melvin universe, a static magnetic/electric field is homogeneous only along the axis of symmetry, and it decreases rapidly in spatial directions orthogonal to the axis. Therefore, it is interesting to consider also the dynamics of black holes in the background space-time with \emph{spatially homogeneous } magnetic/electric field. This background is well known static, conformally flat Bertotti-Robinson space-time with topology AdS${}^2\times \mathbb{S}^2$.

The solution for non-accelerated Schwarzschild black hole immersed into the Bertotti-Robinson magnetic universe was found in \cite{Alekseev-Garcia:1996} (with some correction in \cite{Ortaggio-Astorino:2018}).  Later the solution \cite{Alekseev-Garcia:1996} was used in  \cite{Alekseev:2017} for construction of dynamical exact solution for oscillating ``geodesic'' motion of a Schwarzschild black hole along a magnetic field, where the gravitational field acts as a potential well.

In this paper, another exact dynamical solution for a Reissner-Nordstr$\ddot{o}$m black hole immersed in Bertotti-Robinson space-time with spatially homogeneous electric field and uniformly accelerated by this field is presented.
\vspace{-4ex}
\begin{figure}[h]
\begin{center}
\epsfig{file=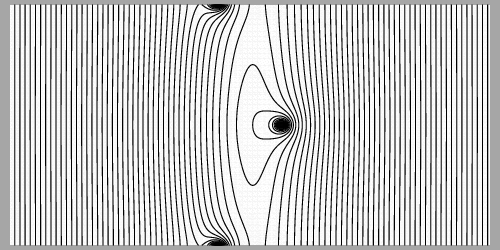,width=2.7in,height=1.3in}
\end{center}
\vspace{-5ex}
\caption{\footnotesize The lines of constant values of metric function $\log (-g_{tt})$ on $(\rho,z)$ plane outside a charged non-rotating black hole accelerated by the external spatially homogeneous (in the absence of a black hole) electric field. The top and bottom boundaries of the picture represent the same axis of symmetry  antipodal to the axis where a black hole is located. The black object on this antipodal axis is a singularity arising due to focussing there of the ``force'' lines.
}
\end{figure}

\section*{Background A$\text{d}$S${}^2\times \mathbb{S}^2$ space-time with spatially homogeneous electric field}
In what follows, we construct the solution of Einstein-Maxwell equations for a charged non-rotating black hole, characterized by metric and electromagnetic 4-potential
\begin{equation}\label{RN-solution}
\begin{array}{l}
ds^2=-H dt^2+H^{-1}dr^2 +r^2(d\theta^2+\sin^2\theta\, d\varphi^2),
\\[0ex]
A_i=\{-\dfrac{e}{r},0,0,0\},\qquad
H=1-\dfrac {2 m}r+\dfrac{e^2}{r^2},
\end{array}
\end{equation}
immersed into the static electrovacuum space-time with spatially homogeneous electric field and geometry  structure AdS${}^2\times \mathbb{S}^2$. Metric and electromagnetic vector-potential of this ``background'' space-time are described by the  electrovacuum solution found by Levi-Civita in 1917, rediscovered later by Bertotti and Robinson (both in 1959) and well known now as the Bertotti - Robinson solution \cite{SKMHH:2003}. For our purposes, we consider this background solution in the form used also in \cite{Alekseev-Garcia:1996}, \cite{Alekseev:2017}:
\begin{equation}\label{BR-solution}
\begin{array}{l}
ds^2=-\left(\cosh \dfrac zb\right)^2\,dt^2+d\rho^2+dz^2+b^2
\left(\sin\dfrac\rho b\right)^2\,d\varphi^2\\[2ex]
A_i=\{\sinh\dfrac z b,0,0,0\},\qquad
E_{\widehat{z}}=\dfrac 1 b,
\end{array}
\end{equation}
where $E_{\widehat{z}}$ is the constant value of electric field directed along the $z$-axis and measured by a local observer.

The constant parameter $b$ possesses a geometrical interpretation: the spatial sections $t=const$ are semi-closed in $\rho$-directions and the radius of this closure is $\rho=\pi b$. Each of two-dimensional sections  $\{t=const,z=const\}$ possesses the geometry of a 2-sphere with the radius $b$. The coordinates on these spheres are $\{\rho,\varphi\}$, where the polar angle is $\rho/b$ and the azimuthal angle is $\varphi$. The circles $\rho=\pi b$ actually are the points "antipodal" to the points $\rho=0$ and constitute the axis of symmetry parallel to and the most distant from the axis $\rho=0$. Due to  symmetries of this space - time,  any of its point can be chosen as the origin of some static frame.

In the limit $b\to\infty$ for finite values of $\rho$ and $z$, the background electric field (\ref{BR-solution}) vanishes and the radius of closure in  $\rho$-direction becomes infinite, i.e. the closure disappears and the background metric (\ref{BR-solution})  becomes flat:
\[
ds^2=-dt^2+d\rho^2+dz^2+\rho^2\,d\varphi^2.
\]
In the static frame (\ref{BR-solution}), the world line of a test particle at rest on the axis $\rho=0$ is not a geodesic provided its location does not coincide with the origin $z=0$. The acceleration of such resting test particle is
\begin{equation} {\bf W} =
\{0,0,W^z,0\}, \qquad W^z = {1\over b}\,\tanh ({z\over b}).
\end{equation}
If the resting test particle of mass $m$ possesses a charge $e$, the Lorentz force, acting on this particle due to its interaction with the background electric field, gives rise to acceleration of this particle equal to $(e/m)E_{\widehat{z}}$. Therefore, the equilibrium position $z=z_o$ of this particle on the $z$-axis is determined by the relation
\begin{equation}\label{TestEquilibrium}
\tanh ({z_o\over b})={e/m}
\end{equation}
It is easy to see that an equlibrium of a charged test particle in the static frame in the space-time (\ref{BR-solution}) is possible only for particles with $\vert e \vert<m$.

\section*{Reissner-Nordstr{\" o}m black hole immersed\\
into A$\text{d}$S${}^2\times \mathbb{S}^2$ electric universe}
The static axisymmetric electrovacuum solution of Einstein-Maxwell equations, described below, represents a nonlinear superposition  of gravitational and electromagnetic fields  of the Reissner-Nordstr{\" o}m black hole (\ref{RN-solution}) and the background fields (\ref{BR-solution}) of Berotti-Robinson space-time. For its construction, the monodromy transform approach with the corresponding linear singular integral equation method, developed  by the author in \cite{Alekseev:1985}-\cite{Alekseev:1992} was used. (Some details of this construction can be found in the last section).  This solution takes the form
\begin{equation}\label{metric}
\begin{array}{l}
ds^2=g_{tt} dt^2+f (d\rho^2+dz^2)+g_{\varphi\varphi} d{\varphi}^2,\\[1ex]
\mathbf{A}=\{A_t,0,0,0\},
\end{array}
\end{equation}
where all components are expressed in terms of bipolar coordinates $\{x_1,y_1,x_2,y_2\}$ and parameters $\{b,\mu,\nu,\sigma,\ell\}$:
\begin{equation}\label{components}
\begin{array}{l}
g_{tt}=-\dfrac{(x_1^2+b^2)(x_2^2-\sigma^2)}{b^2}\left[\dfrac{x_2+\mu y_1+\nu (x_1+\mu y_2)}
{x_2^2-\sigma^2-\nu^2(x_1^2+b^2)}\right]^2,\\[3ex]
g_{\varphi\varphi}=\dfrac{b^2(1-y_2^2)}{x_1^2+b^2}
\left[\dfrac{x_2^2-\sigma^2-\nu^2(x_1^2+b^2)}
{x_2+\mu y_1+\nu(x_1+\mu y_2)}\right]^2,\\[3ex]
A_t=-\dfrac{1}{b}\dfrac{(x_2^2-\sigma^2)(x_1+\mu y_2)\!+\nu(x_1^2+b^2)(x_2+\mu y_1)}{x_2^2-\sigma^2-\nu^2(x_1^2+b^2)},\\[2.5ex]
f=\dfrac{f_o \mathcal{D}^2}{x_2^2-\sigma^2 y_2^2}
\end{array}
\end{equation}
where $f_o$ is an arbitrary positive constant and
\begin{equation}\label{Dfunction}
\begin{array}{l}
\mathcal{D}=[\ell^2+(1-\nu^2)(\mu^2+b^2)][x_2+\mu y_1+\nu(x_1+\mu y_2)]\\[1ex]
-2[\ell\nu+\mu(1-\nu^2)][(\mu^2+b^2)(y_1+\nu y_2)+\ell(x_1-\mu y_2].
\end{array}
\end{equation}
Among four bipolar coordinates, only two ones are independent. The coordinates $\{x_1,y_1\}$ are well adapted to the background geometry and these can be expressed easily in terms of cylindrical coordinates $\{\rho,z\}$, while $\{x_2,y_2\}$ are simply related to black hole coordinates:
\begin{equation}\label{rhoz}
{\left\{\mathbf{\begin{array}{lcll}
x_1=-b\sinh(z/b),&&x_2=(r-m),\\[1ex]
y_1=-\cos(\rho/b),&&y_2=\cos\theta.
\end{array}
}\right.}
\end{equation}
The coordinates $\{x_2,y_2\}$ also can be expressed in terms of the coordinates $\{\rho,z\}$. This can be done using twofold expressions for the known Weyl coordinates $\{\alpha,\beta\}$ for stationary axisymmetric fields:
\begin{equation}\label{alphabeta}
\left\{\begin{array}{l}
\alpha=\sqrt{x_1^2+b^2}\sqrt{1-y_1^2}=\sqrt{x_2^2-\sigma^2}\sqrt{1-y_2^2}\\[1ex]
\beta=\beta_1+x_1 y_1=\beta_2+x_2 y_2,\quad \beta_2-\beta_1=\ell
\end{array}\right.
\end{equation}
From these expressions and from (\ref{rhoz}), we obtain
\[
\left\{\begin{array}{l}
x_2=\dfrac{1}{2}(R_+ +R_-),\\[1ex] y_2=\dfrac{1}{2\sigma}(R_+ -R_-),
\end{array}\right.
\left\{\begin{array}{l}
R_{\pm}=\sqrt{(\beta-\beta_2\pm\sigma)^2+\alpha^2}\\[1ex]
\alpha=b\cosh(z/b)\sin(\rho/b),\\[1ex]
\beta=\beta_1+b\sinh(z/b)\cos(\rho/b),
\end{array}\right.
\]
or $R_{\pm}=\sqrt{(x_1 y_1-\ell\pm\sigma)^2+(x_1^2+b^2)(1-y_1^2)}$.
Alternatively, we can express $(x_1,y_1)$ as functions of $(x_2,y_2)$.

This solution was checked also by its direct substitution into electrovacuum Einstein - Maxwell equations.

\section*{On the parameters of the solution (\ref{metric})--(\ref{alphabeta}).}
The solution, presented in the previous section, depends on four independent real parameters $\{b,\mu,\nu,\ell\}$.
As it was mentioned earlier, the parameter $b$ characterises the background Bertotti-Robinson solution, while $\mu$ and $\nu$ are related to  mass and charge parameters $\{m,e\}$ of the Reissner-Nordstr$\ddot{\text{o}}$m black hole as
\[m=\mu+\dfrac{\nu \ell}{1-\nu},\quad e=\nu \sqrt{(\ell+m)^2+b^2}, \quad \sigma^2=m^2-e^2.
\]
The parameter $\ell$ determines the location of  accelerated black hole on  $z$-axis in the comoving non-inertial frame. Thus, $\ell$ characterises also the black hole acceleration.

\subsection*{Limits of the solution (\ref{metric})--(\ref{alphabeta})}
The solution (\ref{metric})--(\ref{alphabeta}) possesses three important limits.

\underline{\emph{The absence of the external electric field.}} In the limit $b\to \infty$ the external (with respect  to the black hole) electric field vanishes, and the solution (\ref{metric})--(\ref{alphabeta}) transforms into the Reissner-Nordstr{\" o}m black hole solution (\ref{RN-solution}).

\underline{\emph{The absence of the black hole.}} In this limit, $e=m=0$ (and thus, $\sigma$, $\mu$ and $\nu$ vanish), i.e. the black hole is absent and (\ref{metric})--(\ref{alphabeta}) reduces to Bertotti-Robinson solution (\ref{BR-solution}).

\underline{\emph{Schwarzschild  black hole in a Bertotti-Robinson mag-}}
\-{\underline{\emph{netic universe.}} After simple renumeration of coordinates and dual rotation of electromagnetic field, in the limit $e\to 0$ (and $\nu\to 0$), the solution (\ref{metric})--(\ref{alphabeta}) reduces to the solution for a Schwarzschild blak hole immersed into a spatially homogeneous magnetic field \cite{Alekseev-Garcia:1996}.

\section*{The equation of motion of a charged black hole\\ derived from the regularity axis condition}
The equation of motion (\ref{TestEquilibrium}) of a charged test particle in the static electric and gravitational fields, looks like static equilibrium in the co-moving non-inertial frame.

The accelerated motion of a charged black hole in the background space-time with homogeneous electric field also can be described by a static solution. The solution (\ref{metric})--(\ref{alphabeta}), represents a nonlinear superposition of the black hole and external (background) gravitational and electromagnetic fields. In this case, the equation of motion (the condition of equilibrium) of a black hole arises as the condition of regularity of space-time geometry (i.e. the absence of conical singularities) on the parts of axis of symmetry outside the black hole horizon.

In the space-time with metric of the form (\ref{metric}), the point $(\rho=0,z)$ of the $z$-axis  is regular, i.e. a conical singularity is absent there, if the length $\cal L$ and the radius $\cal R$ of a small circle $\{\rho=const,z=const\}$, surrounding the axis, satisfy ${\cal L}\to2 \pi {\cal R}$ for $\rho\to 0$. Given $z$, for small $\rho$, we have
\[{\cal R} \approx\sqrt{f(\rho=0,z)}\, \rho, \quad {\cal L}=2\pi \sqrt{g_{\varphi\varphi}(\rho,z)}.\]
To calculate the limit of $2\pi{\cal R}/{\cal L}$ for $\rho\to 0$, we take into account that, for metrics of the form (\ref{metric}),  $g_{tt} g_{\varphi\varphi} =-\alpha^2$. Using for $\alpha(\rho,z)$ the explicit expression, we obtain
\[P(z)\equiv \lim_{\rho\to 0}\left({{2\pi\cal{R}}}\over {\cal L}\right) = {1\over \cosh(z/b)}\sqrt{-f(0,z) g_{tt}(0,z)}.\]
Calculations for the solution (\ref{metric})--(\ref{alphabeta}) show that locally this limit $P$ is independent of $z$, but it takes different constant values on different parts of the axis outside the black hole horizon. We denote these constants as $P_-$ for $b\,\sinh(z/b)<\ell-\sigma$ and $P_+$ for $b \sinh(z/b)>\ell+\sigma$. For regularity of these parts of the axis we should have
\[P_+=1\quad\text{and}\quad P_- =1.\]
One of these conditions, say $P_+ =1$, can be satisfied by an appropriate choice of an arbitrary positive constant $f_o$ in the expression for $f(\rho,z)$ in (\ref{components}). This choice should be $f_o=1/[\sqrt{(\ell+m)^2+b^2}-e]^4$. In this case, the second condition $P_- = 1$ imposes cirtain relation between the parameters of the solution (\ref{metric})--(\ref{alphabeta}). This relation is:
\begin{equation}\label{Equilibrium}
[(\ell-\sigma)^2+b^2][(\ell+\sigma)^2+b^2]=[\sqrt{(\ell+m)^2+b^2}-e]^4
\end{equation}
with essential parameters $\{b,\ell,m,e\}$, while $\sigma^2=m^2-e^2$.
This is the equation of motion in the form of condition of equilibrium in a non-inertial frame comoving to a charged black hole immersed into the external electric field.  (In particular, the equation (\ref{Equilibrium}) is satisfied for the case, shown in FIG.1 ($b=1$, $m=0.2$, $e=0.14$, $\ell\approx 0.895$).

To make the above statements more explicit, we consider the case of a small black hole, leaving in (\ref{Equilibrium}) only leading linear terms for $m$ and $e$. Then, for equilibrium condition of a small black hole we obtain $m \ell\approx e\sqrt{b^2+\ell^2}$, or in terms of z-coordinate with $\ell=b\sinh[{z_o\over b}]$, in this approximation we have from (\ref{Equilibrium}) the equation
\begin{equation}\label{SmallBH}
\tanh ({z_o\over b})\approx{e/m}
\end{equation}
which coincides in this approximation with the test particle equilibrium condition (\ref{TestEquilibrium}).

\section*{Comments and Conclusions}
The solution (\ref{metric})--(\ref{alphabeta}) of electrovacuum Einstein - Maxwell equations, presented here was constructed using the  {\it monodromy trans\-form approach}, suggested by the author in \cite{Alekseev:1985}-\cite{Alekseev:1992}. In this approach, every local solution with two commuting isometries and vanishing non-dynamical degrees of freedom \cite{Alekseev:2016}, is characterized by the {\it monodromy data} of the corresponding solution of associated linear (spectral) problem, i.e. by two pairs of functions $\{\mathbf{u}_+(w), \mathbf{v}_+(w)\}$ and $\{\mathbf{u}_-(w), \mathbf{v}_-(w)\}$ of a complex (spectral) parameter $w$ which are holomorphic respectively on the parts of the compound cut $L_+\cup L_-$ on $w$-plane. (For vacuum fields, $\mathbf{v}_\pm(w)\equiv 0$).

The components of metric and electromagnetic potential are determined by the solution of a basic system of linear singular integral equations with  kernels  and right hand sides expressed in terms of the monodromy data.

For physically important subclasses of fields, e.g., for stationary axisymmetric fields with regular axis of symmetry, the monodromy data are
{\it analytically adjusted (matched)} and consist of one pair of functions, holomorphic on a common cut $L$: $\mathbf{u}_+(w)=\mathbf{u}_-(w)\equiv\mathbf{u}(w)$ and $\mathbf{v}_+(w)=\mathbf{v}_-(w)\equiv\mathbf{v}(w)$.
With such type of data, the basic system of linear singular integral equations admits explicit solutions for  $\mathbf{u}(w)$ and $\mathbf{v}(w)$ chosen as {\it any rational functions} of $w$, that leads to  hierarhies of  field configurations  with any number of parameters.

A nice feature  of this approach is that the solutions for {\it nonlinear superpositions} of fields may arise from {\it linear superpositions} of  monodromy data of interacting fields.

In particular, for the Reissner-Nordstr$\ddot{o}$m black hole solution (\ref{RN-solution}), the monodromy data possess simple pole structures: $\mathbf{u}(w)=u_1/(w-h)$ and $\mathbf{v}(w)=v_1/(w-h)$ where   $h$ is real and $u_1$, $v_1$ are pure imaginary constants. For  Bertotti-Robinson solution (\ref{BR-solution}), the monodromy data are imaginary constants: $\mathbf{u}(w)=u_0$ and $\mathbf{v}(w)= v_0$.  The solution (\ref{metric})--(\ref{alphabeta}),
describing the nonlinear interaction of a charged non-rotatomg black hole with the external electric field, arises as the solution of the mantioned above linear singular integral equations for the monodromy data, chosen as linear superposition (sum) of the monodromy data of these interacting fields.

An obvious generalization of the solution (\ref{metric})--(\ref{alphabeta}) to the rotating case (Kerr-Newman black hole, immersed into Bertotti-Robinson electric universe), corresponding to the choice of $u_1$, $v_1$ and $h$ as complex constants, is contained in the expressions in the Appendix in \cite{Alekseev-Garcia:1996} and the only problem is to find its more compact form.

From physical point of view, it appears useful to mention that a singular object appeared on the ``antipodal'' axis (FIG. 1) due to semi-closed topology of the background  space-time, is located at very large (cosmological-like) distance from a black hole, provided the value of the background electric field is not unreasonably large. In this case, one may expect that the solution (\ref{metric})--(\ref{alphabeta}) describes  correctly physical properties of fields in rather large space-time region around the black hole.



\end{document}